\documentstyle[aps,preprint]{revtex}

\newcommand{\cl}{{\cal L}}
\begin{document}
%\bibliographystyle{prsty}%
%\draft

\title{ Uncertainty Principle Enhanced  Pairing  Correlations
in Projected Fermi Systems Near Half Filling}

\author{B. Sriram Shastry}
\address{Indian Institute of Science, Bangalore 560012, INDIA}
\maketitle 

%\receipt{}

\date{\today}

\begin{abstract}
We  point out the curious phenomenon of  {\em order by projection} 
in a class of lattice Fermi
systems near half filling.
 Enhanced pairing correlations  of  extended
s-wave Cooper pairs result from  the process of 
projecting out  s-wave  Cooper pairs,    with
negligible effect on the ground state energy.  
The Hubbard model
is a particularly nice example of the above phenomenon,  which
is revealed with the   use of
 rigorous inequalities including  the Uncertainty Principle Inequality
 In addition, we present 
numerical evidence that at half filling,  a related but simplified
model shows ODLRO
of  extended s-wave  Cooper pairs.

\end{abstract}

\begin{flushright}
{Date 19 September 1996, Ver.2}
\end{flushright}
%\narrowtext
%\newpage

There is considerable current interest in the possibility of purely electronic
interactions driven superconductivity as a mechanism to explain the
High $T_c$ superconductors.
While  it is well  known   for the uniform electron
gas that purely 
Coulomb repulsion terms  lead to
superconductivity   in higher angular momentum channels\cite{kl},
 albeit with
very low transition temperatures, here   the search, guided by experiments,
 is predominantly  for single band  models that
display such behavior in the proximity of half filling on  a lattice.
The prototypical example is that of the Hubbard model
 \cite{pwarvb}  \cite{hubmodel},  
 although its tendency (or otherwise) towards superconductivity 
 remains an unsettled issue.

 In this work, we study a class of many body Fermi systems
on a lattice, under the influence of a projection of s-wave Cooper
pairs. Recall that one has an inhibition 
of s-wave ordering within weak coupling
BCS theory for models for on-site
repulsion in addition to the usual 
phononic coupling. In contrast,  we {\em
project out}  s-wave Cooper pairs in the present work. 
The study of  most projected models,   generally
justified  by
their status as ``fixed point'' Hamiltonians in some underlying
scaling theory,  has been 
a rich source of new and interesting models in the field of
correlated fermi systems. A prototype is the Gutzwiller
wavefunction, wherein upon removing double occupancy, 
 effects such as enchanced effective masses
 follow near half filling, and these are crucial in our understanding of
 almost localized fermi liquids\cite{brinkman-rice-gutzwiller}. At half  
 filling  we find  insulating wave functions
with enhanced spin-spin correlations\cite{kaplan-gros-rice-joynt},
that are regarded as typical of Quantum Spin Systems in low dimensions
with $S=1/2$.
At the level of the Hamiltonian, projection leads to interesting new models,
such as the  various limits of the 
Hubbard model,  e.g. large $U$  giving the $t-J$ model,  $U= \infty $ 
giving the
Nagaoka limit, and several examples in single impurity models. 
It seems  worth remarking that  projection
is a  theoretical device that is genuinely strong coupling and
non perturbative, making  it difficult to
 treat with
conventional methods. In the present work 
the
consequences of s-wave projection are  found   by 
combining  a  set of  known inequalities in a novel fashion,
and lead to surprising  insights detailed below.

We  consider the model  defined on a d-dimensional 
hyper cubic lattice with a Hamiltonian
\begin{equation}
H= T + U \sum_i n_{i \uparrow} n_{i \downarrow} + U_s B^\dagger B  
\label{ham}
\end{equation}
where $T$ is the  kinetic energy 
term $\sum_{k,\sigma} \epsilon(k) c^\dagger_\sigma(k) c_\sigma (k) $.
The second term is the  
Hubbard repulsion term (i.e. $U \geq 0$).
We will consider other forms of interaction below, but the argument
is simplest for the Hubbard interaction written above.
We will take the number of sites as ${\cal{L} }$ and denote  the density
of particles by $\rho = N/{\cal{L} }$.  We will also denote 
 $\tilde{H} = H - \mu \hat{N}$, where $\mu$ is the chemical potential and $\hat{N}$ the 
number operator.  
 The third  term is new, with the operator
$B= \sum_i \exp\{i \phi_j\} \;  b_j$ and 
$b_j \equiv c_{j \downarrow} c_{j \uparrow}$.  
If we take $U_s$ to be $O(1/ \cl)$, 
and {\em negative} then this
 term would, in a weak coupling
BCS like theory,  inhibit the formation of s-wave Cooper pairs. 
 The coupling constant $U_s$ is taken of $O(1)$ and {\em positive} 
 in the present
work  and corresponds to {\em projecting out}
the appropriate Cooper pairs, at general fillings. Precisely at half filling,
the influence of the new term is more subtle as noted later in the paper.
 Although various choices of the phase angle $\phi _i$ generate 
different 
 examples,  two of the  interesting ones are 
 ($\alpha$) ~  $\phi_i=0$ leads to a suppression of
pure s-wave superconductivity, and  ($\beta$) ~ $\phi_i = \vec{r}_i \cdot \{ \pi, \pi,.. \} $
suppresses the  so called  eta pairing \cite{yang}.
We will also consider a third possibility ($\gamma$) ~
obtained by setting $B= \sum_k \exp\{i \phi(k)\} \; b(k) $ 
where $b(k) \equiv c_{- k \downarrow} c_{ k \uparrow}$ and 
with an arbitrary function $\phi(k)$ which can be used to vary the relative phases
between different momenta. This last class of operators, however forces us
to the case of $U=0$, in order to obtain any results. The case ($\alpha$) ~
appears to be
 the most interesting physically,  the  others are included
for completeness.

We first note that for lattices that are bipartite,
and where the electronic  hopping only connects unlike sublattices,
we can make a particle hole transformation $c_{i \uparrow} \rightarrow
(-1)^{\theta_i} \; c_{i \uparrow}^\dagger $,  with $\theta_i= 0,1$ for 
the two sublattices, whereby the energy
satisfies $E[U,U_s,\rho] = E[U,U_s, 1- \rho] -   {\cal{L} }(1 - \rho) 
(U + U_s)$.
At half filling ($\rho=1$) the chemical potentials
for adding and subtracting a particle add up 
as: $\mu_+\; + \mu_- \; = U + U_s$. At this  filling, the new 
term $U_s$ plays a  crucial role in allowing 
doubly occupied sites and holes to wander away from each other, and infact
encourages charge fluctuations, whereby the usual Mott insulating state
of the Hubbard model at half filling is heavily discouraged.

We now  use a simple but useful inequality \cite{sw}
\begin{equation}
<\psi_0|\; M^\dagger \;  [\tilde{H}, M]\; |\psi_0> \; \geq \; 0 \label{ineq1}
\end{equation}
where $|\psi_0>$ is the ground state of $\tilde{H} $, and $M$ is an arbitrary
operator. Using $M= B$ we find on using
the important commutator $[B, B^\dagger] = {\cal{L} } - \hat{N} $,
valid in all cases ($\alpha$), ($\beta$) ~ and ($\gamma$),  that
\begin{equation}
<B^\dagger A> \geq \{ U_s({\cal{L} } - N + 2) - 2 \mu + U \} < B^\dagger B>. 
\label{ineq2}
\end{equation}
In the case of ($\gamma$), the above Inequality holds only with $U=0$. Note that the LHS of above is  forced to be real and  to be positive
from the Inequality. We  denote ground state averages by angular brackets as above, and
the operator $A$ is given by $A= [T,B] $. For the two cases of the 
phase $\phi_i$ in Eq(\ref{ham}) , ($\alpha$) $A= - 2 \sum_k \epsilon(k) b(k)$,
 and ($\beta$)
$A= - \sum \{ \epsilon(\vec{k}) + \epsilon(\vec{k} + \vec{\Pi}) \} \; b(k)$ with
$\vec{\Pi}= \{ \pi, \pi,..\}$. In  the  popular case of nearest neighbour 
hopping on the hypercubic lattice, ($\alpha$)  corresponds to the extended s-wave
pairing operator, whereas  ($\beta$) gives zero. A non zero result is obtained in
the latter case only when the hopping connects sites on the same sublattice.
In  two dimensions, for example, with $\epsilon(k) = - 2 t ( \cos(k_x) +
\cos(k_y) ) - 2 t' \cos(k_x) \cos(k_y) $, we find $A = 4 t'
\sum(\cos(k_x) \cos(k_y) ) \; b(k)$.

We now use the Cauchy-Schwartz inequality to  bound the LHS of
Inequality(\ref{ineq2}) as 
\begin{equation}
<A^\dagger A> \; <B^\dagger B>\; \geq \; <B^\dagger A>^2 .\label{ineq3}
\end{equation}  
 Combining Ineqs(\ref{ineq2} ,\ref{ineq3} )
we find
\begin{equation}
<A^\dagger A> \; \geq \; \{ U_s({\cal{L} } - N + 2) - 2 \mu + U \}^2 \; < B^\dagger B>. \label{ineq4}
\end{equation}
Again note that in case of ($\gamma$),  Ineq(\ref{ineq4}) is valid only with $U=0$.
We note that in the RHS of Ineq(\ref{ineq4}) the prefactor is of the $O({\cal{L} }^2)$
 provided we are at a thermodynamic filling $\rho < 1$. Exactly at
half filling the inequality  is less useful.
 At any filling $\rho <1$,  we can deduce  that
$<B^\dagger B>$ is very small.  In fact we will show that it
is $o(\cl)$ rather than $O(\cl)$.
If it were of the $O({\cal{L} })$, ( as indeed it
is in the ground state of the free fermi gas), then $<A^\dagger A>$
has to  exceed
a trivial upper bound of the $O({\cal{L} }^2)$\cite{boundona}. 
If $<B^\dagger B>$
is of the $O(1)$ then we have two consequences that are mutually incompatible
(at least when $U=0$). To see this, assume that 
$<B^\dagger B>$ is of the $O(1)$
and so we find from the Feynman Hellman Theorem
\begin{eqnarray}
E[U,U_s,\rho] & = & E[U,0,\rho] + \int_0^{U_s} \; dU_s' 
\; <B^\dagger B>_{U_s'}\\
& = & E[U,0,\rho] + o(\cl). \label{energyunchanged}
\end{eqnarray}
The other consequence of Ineq(\ref{ineq4})  is that $<A^\dagger A> \sim O({\cal{L} }^2)$, i.e. we
have Long Ranged Order (ODLRO) in the operator $A$\cite{yanglro}.  This is possible only if
the energy increases by terms of the $O({\cal{L} })$, at least in the case
when $U=0$ as is seen from a diagonalization of a bilinear Hamiltonian
adding the kinetic energy $T$ and $A$ with  
coefficients of the $O(1)$\cite{remark1}. A  
consistent possibility\cite{remark2} is
\begin{equation}
<B^\dagger B>  =  O(1/{\cal{L} })  \; \mbox{and} \;
 <A^\dagger A>  =  O({\cal{L} }),       \label{finbounds}
\end{equation}
along with Eq(\ref{energyunchanged}).
One immediate consequence of this result is that
the energy per site  of the model in Eq(\ref{ham}) 
$ Lim_{\cl \rightarrow \infty} E[U,U_s,\rho]/ \cl $ is identical to that of
the pure Hubbard model ( i.e. $U_s =0$) at {\em all U} or  filling $\rho 
\neq 1$.

Another important consequence is that the
chemical potential is unchanged by $U_s$ until we reach half filling ($\mu=
\partial E / \partial N $) and therefore the
 compressibility
is unchanged by $U_s$
(since  $1/\kappa = N \rho (\partial \mu /\partial N )_{\cal{L} }$). At precisely half filling,
 the chemical potential jumps and the compressibility vanishes.
The value of   $\mu$  at half filling
for the case of bipartite symmetry  was given above as $(U + U_s)/2$.

We see that the suppression of a correlation of the type $<B^\dagger B>$
occurs  with remarkable efficiency through Ineq(\ref{finbounds}). When
we recall that $<[B,B^\dagger]> = {\cal{L} } ( 1 - \rho)$, it is seen that the
fluctuations of $B + B^\dagger$ in the ground state diminish
on approaching half filling, 
i.e. $< (B + B^\dagger)^2/{\cal{L} }>  = 1- \rho$.
This immediately suggests that ``conjugate variables''  in the sense
of the Uncertainty Principle, should exhibit enhancements by similar factors.
The following form of the Uncertainty Principle is most useful, for
any two operators $ a$ and $b$  ( such that $<a^2> =0=<b^2>$ and $[a,b]=0$)
we have \cite{remark3}
\begin{equation}
<a^\dagger a + a a ^\dagger > \; < b^\dagger b + b b^\dagger> \; \geq \;
|< [ a^\dagger,b] >|^2. \label{uncprin}
\end{equation}
We now use this with $a \rightarrow A$ and $b \rightarrow B$, and also
the results $[A,B^\dagger]=  2 T$ and $ [A, A^\dagger] \equiv \chi_A=4 \sum_k
\epsilon(k)^2 (1 - \sum_\sigma c^\dagger_\sigma (k) c_\sigma(k) )$
to find
\begin{equation}
<A^\dagger A> \; \geq   
\frac{ 2 |<T>|^2}{{\cal{L} } (1- \rho)} \;  
-\frac{<\chi_A>}{2}. \label{uncerta}
\end{equation}
Both terms of the RHS of the above Inequality are of the $O({\cal{L} })$, and the second term remains
bounded as we approach half filling, infact vanishing
for the case of a symmetric band around zero energy.
 This implies that the first term dominates
and hence we conclude that the correlation function $<A^\dagger A>$ grows
without limit as half filling is approached.  We should remark that  any
operator of the type $[T,[T,..[T,B]..]$ in the place of $A$ would end up 
having similar enhancements in its correlations, since it
 would be a bilinear in the $c^\dagger$ and have similar commutation with $B$

We next consider other kinds of interactions, 
different from the Hubbard model\cite{rmotherorders}. In this case, we can still use
Ineq(\ref{ineq1}) and also Ineq(\ref{ineq3}) to
find in place of Ineq(\ref{ineq4}), 
\begin{equation}
<F^\dagger F> \geq \{U_s(\cl -N + 2) - 2 \mu \}^2 \; \; <B^\dagger B>,\label{fineq}
\end{equation}
where $F= A + C $  with $C= - [B, V_{int}]$, so that 
$F= [H ,B] $.  The norm on the LHS of Ineq(\ref{fineq}) can be bounded by  
the triangle inequality as $<F^\dagger F> \leq [\sqrt{<A^\dagger A>} +
\sqrt{<C^\dagger C>} ]^2 $ and hence we need, in addition to the 
previous estimates, one of $<C^\dagger C>$. This of course depends
upon the  nature of the two particle interaction, and has to be examined
for each model separately. However, for ``generic'' repulsive 
short ranged models,
it seems clear that this object, like $<A^\dagger A>$ should be bounded
from above by a  number of $O(\cl^2)$. With this assumption, the
remaining argument goes exactly as in the case of the Hubbard model,
and we again conclude that $<B^\dagger B>$ is at least
as small as $o(\cl)$, and in fact probably $O(1/ \cl)$, and that
the ground state energy is as in Eq(\ref{energyunchanged}). 
The uncertainty relation  Ineq(\ref{uncerta}) needs only the fairkly weak 
first condition $<B^\dagger B> \sim o(\cl)$, and hence
we conclude that the mechanism of {\em order by projection} works
for generic short ranged repulsive models near half filling.

We have thus found enhanced correlations as we approach half filling,
and by continuity, we may expect ODLRO
in the operator $A$. The inequalities given above do not 
constrain correlations sufficiently, and we turn to other methods. Before
doing that, we introduce a simpler version of the models above, namely
\begin{equation}
\tilde{H}_s = \sum_{n=1}^L (\epsilon_n - \mu) (\sigma^z_n+1) + U_s \sum_{n,m=1}^{L} \sigma^+_n \sigma^-_m, \label{hs} 
\end{equation}
where $\sigma^z$ etc are the usual Pauli matrices, and $\epsilon_n$ are an
ascending set of energies. This model is intimately related to the 
$U=0$ version of our starting problem Eq(\ref{ham}), using the
pseudo spin representation $\sigma^z_j +1 = \sum_\sigma n_\sigma(k_j)$
and $\sigma_j^+= c_\uparrow^\dagger(k_j) c_\downarrow^\dagger(-k_j)$,
in the subspace where both $(k,\uparrow)$ and $(-k,\downarrow)$ 
are simultaneously present or absent. The Hamiltonian Eq(\ref{ham})
commutes with the operator $\nu= \sum_k n_\uparrow(k) n_\downarrow(-k)$,
and its operation is identical to that of $H_s$ provided we
specialize to various sectors labeled by the eigenvalues of
$\nu$ ~($0 \leq \; <\nu> \; \leq N/2 $), and further
 choose
appropriate degeneracies for the  energies.
 We simplify by choosing our energies in
$H_s$ above to be  non degenerate, and pick  them to be
$\epsilon_n = \{ n - (L+1)/2 \} / ( L-1)$ so that the band
is symmetric about zero and the bandwidth is unity.
Each up spin corresponds to two ( fermi) particles of the original problem.
 The filling in this problem is clearly $\rho =N/L$ with
 $\hat{N}= \sum_j ( \sigma^z(j)+1)$ .
The chemical potential at half filling is $U_s/2$ by particle hole symmetry.

The model can also be viewed as a 
lattice  of N/2 hard core  particles  sitting in a constant
electric field  that tries to localize them in regions of low potential,
and an infinite ranged hopping that tries to delocalize them.
The results proved for the starting Eq(\ref{ham})
namely Ineqs(\ref{ineq4},\ref{uncerta},\ref{finbounds}), are equally true in
this one dimensional spin model, provided we identify  $B= \sum_j \sigma^-_j$
and $A= - 2 \sum_j  \epsilon_j \sigma^-_j$.  Away from half filling,
i.e. when $\sigma^z_{tot} \neq 0$, we see that even in the limit
of large $U_s$, there is a large number of states, infact states
with $S^{tot}=  L(1-\rho)/2$ and $S^z_{tot}= - L(1-\rho)/2$ i.e. highest weight
states of the rotation group, which have a null eigenvalue of the hopping
term $U_s \sum \sigma^+_n \sigma^-_m$. The Zeeman energy term has non zero
matrix elements {\it within this manifold}. In the case of half filling
$\rho=1$, the Zeeman term necessarily connects singlet states with triplets
and hence the energy is unable to escape the influence of $U_s$. At half filling
and for large $U_s$, we can  use degenerate perturbation theory to
find an effective Hamiltonian to lowest order in $1/U_s$. To do this
we consider the action of $H_s$ in Eq(\ref{hs}) on the space of $^{L}C_{L/2}/(L/2+1)$ singlets spanned, for example by the Non Crossing
Rumer Diagrams\cite{rumer}. A typical  non orthogonal state is given by
$\psi_P= [P_1,P_2]_- \ldots [P_{L-1},P_L]_-$ where $P$ is one of the permutations
of the set $\{ 1,2,\ldots,L\}$  giving a Non Crossing Rumer Diagram, and
$[i,j]_\mp = (\alpha_i \beta_j \mp \beta_i \alpha_j)/\sqrt{2}$ is a 
singlet (triplet) with $s^z=0$.
 The action of the  operator Eq(\ref{hs}) can 
be projected into this subspace,    by using the relation
$[1,2]_+ [3,4]_+ = \frac{1}{3} ( 1- 2 \Pi_{13})[1,2]_-[3,4]_- \; + \psi^{quintet}$
with $\Pi_{ij}$ the permutation operator,  and leads to the following Quantum Dimer problem:
\begin{eqnarray}
 & &H_{qd} \psi_P = \frac{-1}{2 U_s} \sum_j (\epsilon_{P_j}- \epsilon_{P_{j+1}})^2
\; \psi_P  -  \nonumber \\ 
& &\frac{1}{3 U_s} \sum_{j+1 < k}  (\epsilon_{P_j}- \epsilon_{P_{j+1}})
(\epsilon_{P_k}- \epsilon_{P_{k+1}}) \{ 2 \Pi_{P_j P_k} -1\} \; \psi_P. \label{qdimer}
\end{eqnarray}
This model is quite non trivial
to work with, but does  reveal 
that the diagonal terms favour singlet bonds that connect the largest energy
separations, and the mixing terms oblige us to take non trivial
linear combinations in this space. 

We  study the interesting half filled limit by studying the sector
$\sigma^z_{total}=0$ of Eq(\ref{hs}) directly. We 
diagonalized the problem numerically for chains
of  length  up to
14, and studied the ground state energy as well as the 
correlation function $<A^\dagger A>$. It is clear that 
a non zero extrapolation of
$\Gamma$ to a number of the $O(1)$ would imply ODLRO in the $A$ field.
In the 
figure  we plot the parameter 
$ \Gamma =  \frac{1}{\L} <A^\dagger A> /  < A^\dagger A>_{non} $
for three values of $U_s$  ($=2,4,8$).
The data seems to be consistent with this hypothesis, and fits well
to $\Gamma = \Gamma_{\infty}+ |a|/L \pm |b| /L^2$ , with non zero $\Gamma_\infty$.
 In the inset of the  figure,  the  ground state energy per site
is plotted for the same
three values of $U_s$ against $1/L$,  showing that the energy 
{\em does}  depend on the coupling at half filling, implying that
the $U_s$ term 
cannot be viewed as a projection at this particular filling. The 
dependence
is consistent with finite sized scaling with a  
form $E/L = e_\infty+ |a| /L - |b|/L^2 + O(1/L^3)$.

 At half filling, the new model Eq(\ref{ham}) is almost certainly non-insulating,
and  likely to be  superconducting  in a complementary pairing state. 
The presence of hopping terms for the doubly occupied sites makes
their number density nonzero, unlike in the pure Hubbard model. By continuity
in filling, we expect the pairing correlations to be divergent for any $U$.
Our numerical results for the reduced model, the spin model of Eq(\ref{hs})
are consistent with ODLRO at half filling. It is not, however, straightforward
to write down a mean field theory that captures the correct ordering 
in the model, since the Hamiltonian does not contain explicit terms that 
favour {\em any} kind of ordering, these are generated by  
the dynamics rather indirectly.

In summary we have seen that the  effect of projecting out s-wave Cooper
pairs in a  class of  Fermi systems leads  to surprising
results.
 The ground state of the projected model may be viewed
as being essentially degenerate with that of the original model
and yet the  extended s-wave pairing correlations are  hugely enhanced
near half filling. 
This effect,  namely {\em order by projection}
requires a lattice Fermi system near half filling  
to occur, and has no natural
counterpart in continuum Fermi
systems. In this regard, as well as in the form of the enhancements
$1/ (1- \rho)$,
it resembles the results of the almost localized fermi systems
\cite{brinkman-rice-gutzwiller}.

\begin{center}
{\bf Figure Captions}
\end{center}
\begin{itemize}
\item Long Ranged Order parameter $\Gamma \equiv  \frac{1}{L} 
{<A^\dagger A>}/{<A^\dagger A>_{non} }$ versus
$1/L$ for $U_s=2,4,8$ (bottom to top) 
chains of length $4,6,..,14$ at $\rho=1$. The inset shows the Ground State 
Energy per site for the same values of $U_s$ versus $1/L$.
(bottom to top).
\end{itemize}
%\widetext

\begin{references}
\bibitem{kl} W. Kohn and J. M. Luttinger, Phys. Rev. Letts. {\bf 54} 524 (1965).
\bibitem{pwarvb} P. W. Anderson, Science {\bf 235} 1196 (1987). A selection
of other theoretical approaches may be found in e.g. ``High Temperature
Superconductivity"", Proceedings Los Alamos Symposium ed. K. Bedell et.al.
Addison Wesley, New York (1990).
\bibitem{hubmodel} ``The Hubbard Model: a reprint Volume'', ed. A. Montorosi,
World Scientific ( Singapore 1992).
\bibitem{brinkman-rice-gutzwiller} M. Gutzwiller,
Phys. Rev. {\bf 137}, A1726 (1965),
W. Brinkman and T. M. Rice, Phys. Rev {\bf B2},4302 (1970), D. Vollhardt
Rev. Mod. Phys. {\bf 56} 99 (1984).
\bibitem{kaplan-gros-rice-joynt} T.Kaplan, P.Horsch and P.Fulde, Phys. Rev. Letts. 
{\bf 49}, 889(1982);
 C.Gros, R.Joynt and T.M.Rice, Phys. Rev.{\bf B 36}, 381(1987);
 F.Gebhardt and D.Vollhardt, Phys. Rev. Letts. {\bf 59}, 1472(1987).
\bibitem{yang} C. N. Yang, Phys. Rev. Letts. {\bf 63} 2144 (1989),
 C. N. Yang and S.C.Zhang, Mod. Phys. Letts. {\bf B4} 759 (1990).
\bibitem{sw} K. Sawada and C. S. Warke, Phys. Rev. {\bf 133} A1252 (1964). 
The inequality
is readily proved by inserting a complete set of energy eigenfunctions.
\bibitem{boundona}
The bound ~is ~obtained by writing
$<A^\dagger A>  \leq 4  \sum_{k,k'}|\epsilon(k)  \epsilon(k') 
 < b^\dagger(k') b(k) > |  \leq \;   4 \sum_{k,k'}|\epsilon(k) \;
\epsilon(k')| \sim O({\cal{L} }^2)$.
\bibitem{yanglro} C. N. Yang, Rev. Mod. Phys. {\bf 34} 694 ( 1962).
\bibitem{remark1} In the case of $U \neq 0$,  provided that $A$
does not already have Long Ranged Order in the absence of the $U_s$,  then the same is  expected to be true, but harder to prove rigorously.
\bibitem{remark2} If $<B^\dagger B> \sim O(\cl^{1- \sigma})$ with
$ 1> \sigma > 0$ then we find from Ineq(\ref{ineq4}) that $<A^\dagger A>$
has quasi ODLRO, i.e.  is almost ordered, unlike in a normal Fermi Liquid.
\bibitem{remark3}  L. Pitaesvski and S. Stringari, J. Low. Temp. Phys. {\bf 85} 377 (1991).
\bibitem{rmotherorders} For other kinds of projected order, the inequalities are harder to interpret e.g.
if $B= \sum_{i,j} [\nu]_{i,j} c_{i \downarrow} c_{j \uparrow}$
with an off diagonal matrix $[\nu]$,
 then the analog of Ineq(\ref{ineq2}) contains,
as a coefficient of $U_s$ the factor $(\cl-N+2)[\nu^2]_{i,i} + <\chi|\phi|\chi>$, where
$|\chi> \equiv B|\psi_0>$ and $\phi=  - \sum_{i,j} [\nu^2]_{i,j} c^\dagger_{i \sigma} c_{j \sigma}$,
and hence there is the
 possibility of cancellation of the term of $O(\cl)$.
\bibitem{rumer} G. Rumer, Nachr. d. Ges. d. Wiss. zu Gottingen, M.P. Klasse,
337 (1932), L. Hulth\'{e}n, Ark. Mat. Astr. Fys., {\bf 26 A}, 1 ( 1938). A
R. Saito , J of Phys. Soc. Japan {\bf 59} 482 (1990).
\end{references}
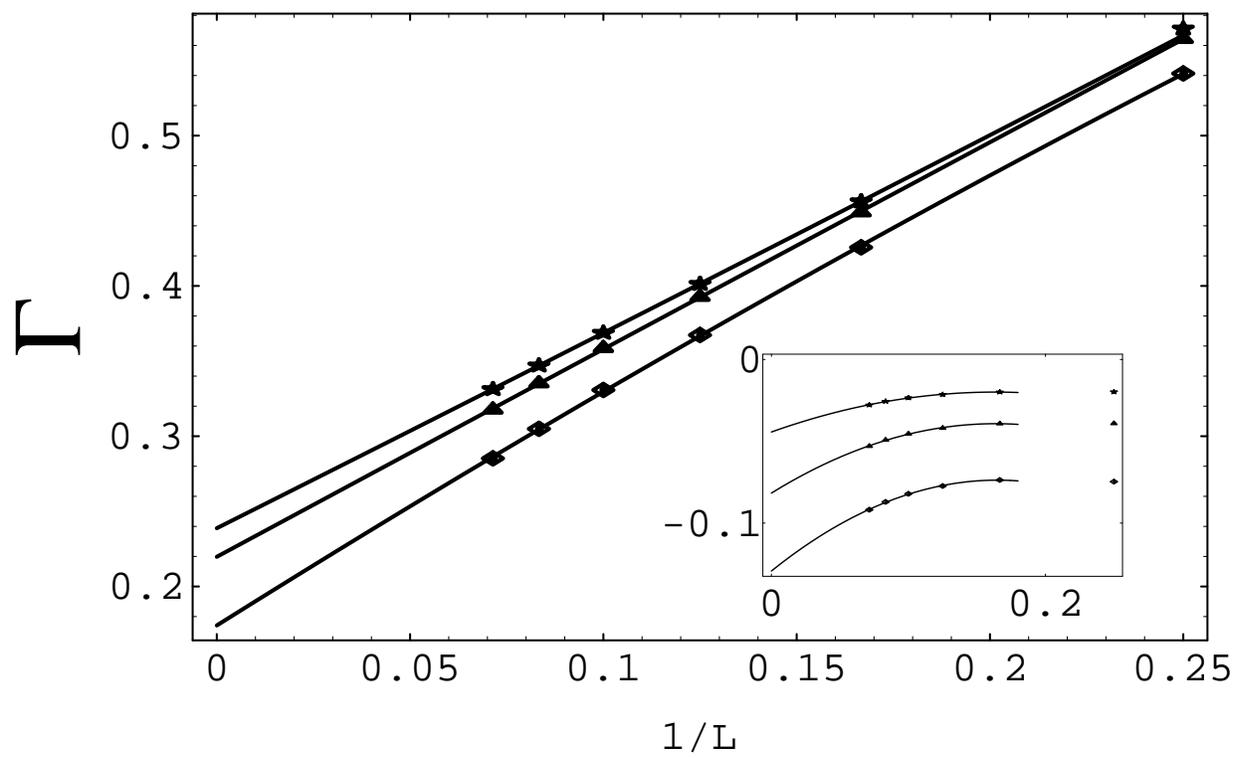
\end{document}